\documentclass[aps,prb,twocolumn,superscriptaddress,floatfix,showpacs,10pt,letterpaper,noshowpacs]{revtex4}

\pdfoutput=1

\newcommand{\RZ}{\R/\Z}

\begin{document}

\begin{titlepage}

\title{
A lattice non-perturbative definition of an $SO(10)$ chiral gauge theory\\
and its induced standard model
}

\author{Xiao-Gang Wen}
\affiliation{Perimeter Institute for Theoretical Physics, Waterloo, Ontario, N2L 2Y5 Canada}
\affiliation{Department of Physics, Massachusetts Institute of Technology, Cambridge, Massachusetts 02139, USA}

\begin{abstract}
The standard model is a chiral gauge theory where the gauge fields couple to
the right-hand and the left-hand fermions differently.  The standard model is
defined perturbatively and describes all elementary particles (except
gravitons) very well.  However, for a long time,  we do not know if we can have
a non-perturbative definition of standard model as a Hamiltonian quantum
mechanical theory.  In this paper, we propose a way to give a modified standard
model (with 48 two-component Weyl fermions) a non-perturbative definition by
embedding the modified standard model into a $SO(10)$ chiral gauge theory.  We
show that the $SO(10)$ chiral gauge theory can be put on a lattice (a 3D
spatial lattice with a continuous time) if we allow fermions to interact.  Such
a non-perturbatively defined standard model is a Hamiltonian quantum theory
with a finite-dimensional Hilbert space for a finite space volume.  More
generally, using the defining connection between gauge anomalies and the
symmetry-protected topological orders, one can show that any truly anomaly-free
chiral gauge theory can be non-perturbatively defined by putting it on a
lattice in the same dimension.

\end{abstract}


\maketitle

\end{titlepage}

%{\small \setcounter{tocdepth}{1} \tableofcontents }

\noindent
\textbf{Introduction}:
The $U(1) \times SU(2)\times SU(3)$  standard
model\cite{G6179,W6764,SW6468,G6267,Z6424,FG7235} is the theory which is
believed to describe all elementary particles (except gravitons) in nature.
However, the standard model was defined only perturbatively initially, via the
perturbative expansion of the gauge coupling constant.  Even though the
perturbative expansion is known to diverge, if we only keep the first a few
orders of the perturbative expansion, the standard model produces results that
compare very well with experiments.

So the ``perturbatively defined standard model'' (keeping only first a few
orders of the perturbation) is a theory of nature.  However,  the
``perturbatively defined standard model'' is certainly not a ``Hamiltonian
quantum theory'' (by keeping only  a few orders of the perturbation, the
probability may not even be conserved).  ``Hamiltonian quantum theory'' is a
quantum theory with\\
(1) a finite dimensional Hilbert space for a finite space volume,\\
(2) a local Hamiltonian operator for the time evolution,\\
(3) operators to describe the physical quantities.\\
So far, we do not know if there is a non-perturbatively defined standard model
which is a Hamiltonian quantum theory.  In this paper, we like to address this
issue.  We will propose a way to obtain a non-perturbative definition of the
standard model that defines the standard model as a Hamiltonian quantum theory.

Defining standard model non-perturbatively  is a well-known long standing
problem, which is referred generally as chiral-fermion/chiral-gauge problem.
There are many previous researches that try to solve this general problem.
There are lattice gauge theory approaches,\cite{K7959} which fail since they
cannot reproduce chiral couplings between the gauge field and the fermions.
There are domain-wall fermion approaches.\cite{K9242,S9390} But the gauge
fields in the domain-wall fermion approaches propagate in one-higher dimension:
4+1D.  There are also overlap-fermion
approaches.\cite{NN9362,NN9474,L9995,N0003,S9947,L0128} However, the
path-integral in overlap-fermion approaches may not describe a Hamiltonian
quantum theory (for example, the total Hilbert space in the overlap-fermion
approaches, if exist, may not have a finite dimension, even for a space-lattice
of a finite size). There are also the mirror fermion approach used in
\Ref{EP8679,M9259,BMP0628,GP0776}, which start with a lattice model containing
chiral fermions \emph{and} a chiral conjugated mirror sector both coupled to
gauge theory.  Then, one tries to include proper direct interaction or boson
mediated interactions\cite{S8456,S8631} between fermions hoping to gap out the
mirror sector only without breaking the gauge symmetry (for more details, see
Appendix \ref{EP}).  However, later work either fail to demonstrate
\cite{GPR9396,L9418,CGP1247} or argue that it is almost impossible to gap out
the mirror sector without  breaking the gauge symmetry in some mirror fermion
models.\cite{BD9216} Some of those negative results are based on some
particular choices of fermion interactions for some particular chiral gauge
theories.

In \Ref{W1303}, a deeper understanding of gauge anomalies and gravitational
anomalies is obtained through symmetry-protected topological (SPT) orders and
topological orders in one-higher dimensions.  This leads to a particular way to
construct mirror fermion models and a  particular way to construct interactions
between fermions.  Such a construction leads to a complete solution of
chiral-fermion/chiral-gauge problem: \frm{By definition, any chiral
	fermion/boson theory can be non-perturbatively defined as a low energy
	effective theory of a lattice theory of finite degrees of freedom per
site by including proper interactions between fermions/bosons, provided that the
chiral fermion/boson theory is free of \emph{all} anomalies.} In other words,
the lattice gauge theory approach actually works (\ie can be used to define any
truly-anomaly-free chiral-gauge theories), provided that we include proper
interactions between fermions/bosons.  In \Ref{WW1380}, we show that a 1+1D $U(1)$
chiral fermion/boson theory is free of \emph{all} the $U(1)$ gauge anomalies if
it is free of the Adler-Bell-Jackiw (ABJ) $U(1)$ gauge
anomaly.\cite{A6926,BJ6947}  We further show that all the ABJ-anomaly-free 1+1D
$U(1)$ chiral fermion/boson theories can be non-perturbatively defined by
$U(1)$ lattice theories.  Even chiral fermion/boson theories with certain
global anomalies can be defined on lattice if the  global anomalies allow a
fully gapped low energy dynamics.\cite{W1303}

However, in general, we do not know how to check if a
chiral gauge theory is free of \emph{all} anomalies.\cite{W1375}
So the above result is hard to use.  To address this problem, in this paper, we
argue that \frm{\textbf{Conjecture:} A
chiral fermion theory in $d$-dimensional space-time with a gauge group $G$ is
free of all gauge and gravitational anomalies if (1) there exist (possibly
symmetry breaking) mass terms that make all the fermions massive, and (2)
$\pi_n(G/G_\text{grnd})=0$ for $n\leq d+1$, where $G_\text{grnd}$ is the
unbroken symmetry group.} Such a conjecture allows us to show that \frm{the
$SO(10)$ chiral fermion theory in the $SO(10)$ grand unification\cite{FM7593}
can appear as a low energy effective theory of a lattice gauge model in 3D
space with a continuous time, which has a finite number of degrees of freedom
per site.} In other words, the lattice-gauge/mirror-fermion approach works for the $SO(10)$
chiral fermion theory.  Following \Ref{W1303}, we propose a way to design a
proper fermion interactions that can gap out the mirror sector only, without
breaking the $SO(10)$ gauge symmetry.

By embedding the modified standard model into the $SO(10)$ grand unification
model,\cite{FM7593,GG7438} the above non-perturbatively defined $SO(10)$ chiral
fermion theory gives a non-perturbative definition of a modified standard
model.  Compare to the standard model, the modified standard model contains a
total of 48 two-component Weyl fermions (one extra neutrino for each family).

In the rest of this paper, we will first give a brief review of the connection
between gauge anomalies and SPT orders.\cite{W1303}  Next we will describe a
particular construction that gives a general non-perturbative definition of all
weak-coupling  chiral gauge theories that are free of all anomalies. Then, as a
key result, we will show that, using such a construction, the modified standard
model (with 48 two-component Weyl fermions) and its corresponding $SO(10)$
chiral gauge theory can be defined as a 3D lattice $SO(10)$ gauge model with a
continuous time (\ie the low energy effective theory of the lattice $SO(10)$
gauge model is the modified standard model).

\noindent
\textbf{Gauge anomalies and SPT orders in one-higher dimension}:
To understand gauge anomalies in weak-coupling gauge theories, we can take the
zero coupling limit.  In this limit, the gauge theory become a theory with a
global symmetry described by group $G$.  Through such a limit, we find that we
can gain a systematic understanding of gauge anomalies through SPT
states.\cite{W1303}

What are SPT states?  SPT states\cite{GW0931,PBT0959} are short-range entangled
states \cite{CGW1038} with an on-site symmetry\cite{CLW1141,CGL1172,CGL1204}
described by a symmetry group $G$.  It was shown that different SPT states in
$(d+1)$-dimensional space-time are classified by  group cohomology class
$\cH^{d+1}(G,\RZ)$.\cite{CLW1141,CGL1172,CGL1204} The SPT states have very
special low energy boundary effective theories, where the symmetry $G$ in the
bulk is realized as a \emph{non-on-site} symmetry on the boundary.  (We will
also refer non-on-site symmetry as anomalous symmetry.) It turns out that the
non-on-site symmetry (or the  anomalous symmetry) on the boundary is not
``gaugable''.  If we try to gauge the non-on-site symmetry, we will get an
anomalous gauge theory, as demonstrated in
\Ref{CGL1172,LV1219,LW1224,CW1217,SL1204} for $G=U(1),SU(2)$.  This relation
between SPT states and gauge anomalies on the boundary of the SPT states allows
us to obtain a systematic understanding of gauge anomalies via the SPT states
in one-higher dimension.  In particular, one can use different elements in
group cohomology class $\cH^{d+1}(G,\RZ)$ to classify (at least partially)
different bosonic gauge anomalies for gauge group $G$ in $d$-dimensional
space-time. This result applies for both continuous and discrete gauge groups.
The free part of $\cH^{d+1}(G,\RZ)$, Free$[\cH^{d+1}(G,\RZ)]$, classifies the
well known ABJ  anomalies\cite{A6926,BJ6947} for both bosonic and
fermionic systems.  The torsion part of $\cH^{d+1}(G,\RZ)$ correspond to new
types of gauge anomalies beyond the Adler-Bell-Jackiw anomalies (which will be
called nonABJ gauge anomalies).\cite{W8224}

Note that the global symmetry of the bulk SPT state in the $d+1$ dimensional
space-time is on-site and gaugable. If we gauge the global symmetry, we obtain a
non-perturbative definition of an anomalous gauge theory in  $d$ dimensional
space-time.  The $d$ dimensional  anomalous gauge theory is defined as the
boundary theory of the $d+1$ dimensional (gauged) SPT state.  We see that an
anomalous gauge theory is not well defined in the same dimension, but it can be
defined as the boundary theory of a (gauged) SPT state in one-higher dimension.
In the next section, we will show that an anomaly-free chiral gauge theory can
always be defined as a lattice gauge theory in the same dimension.

\noindent
\textbf{A non-perturbative definition of anomaly-free chiral gauge theories}:
Motivated by the connection between the chiral gauge theories in
$d$-dimensional space-time and the SPT states in $(d+1)$-dimensional
space-time, we like to show that one can give a non-perturbative definition
for any anomaly-free chiral gauge theories.

\begin{figure}[tb]
\begin{center}
\includegraphics[scale=0.42]{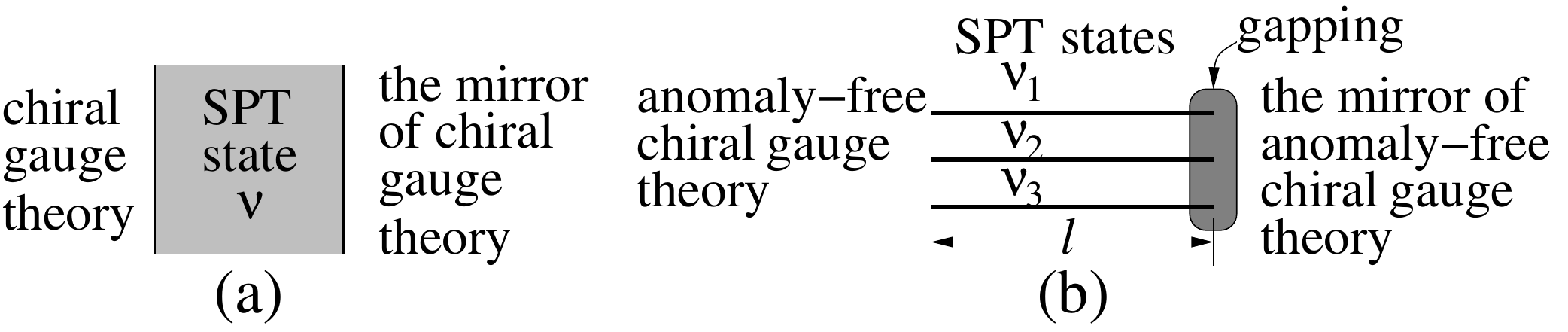}
\end{center}
%Fig. 2
\caption{(a) A SPT state described by a cocycle $\nu \in
\cH^{d+1}(G,\R/\Z)$ in $(d+1)$-dimensional space-time.  After ``gauging'' the
on-site symmetry $G$, we get a bosonic chiral gauge theory on one boundary and
the ``mirror'' of the  bosonic chiral gauge theory on the other boundary.  (b)
A stacking of a few SPT states in $(d+1)$-dimensional space-time described by
cocycles $\nu_i$.  If $\sum_i \nu_i=0$, then after ``gauging'' the on-site
symmetry $G$, we get a anomaly-free chiral gauge theory on one boundary. We
also get the ``mirror'' of the  anomaly-free chiral gauge theory on the other
boundary, which can be gapped without breaking the ``gauge symmetry''.  }
\label{chgauge}
\end{figure}

Let us start with a SPT state in $(d+1)$-dimensional space-time with an on-site
symmetry $G$ (see Fig. \ref{chgauge}a). We assume that the SPT state is
described by a cocycle $\nu \in \cH^{d+1}(G,\R/\Z)$.  On the $d$-dimensional
boundary, the low energy effective theory will have a non-on-site symmetry (\ie
an anomalous symmetry) $G$.  Here we will assume that the $d$-dimensional
boundary excitations are gapless.  After ``gauging'' the on-site symmetry $G$
in the $(d+1)$-dimensional bulk, we get a chiral gauge theory on the
$d$-dimensional boundary whose anomaly is described by the cocycle $\nu$.

Then let us consider a stacking of a few bosonic SPT states in
$(d+1)$-dimensional space-time described by cocycles $\nu_i \in
\cH^{d+1}(G,\R/\Z)$ where the interaction between the SPT states are weak (see
Fig. \ref{chgauge}b).  We also assume that $\sum_i \nu_i=0$.  In this case, if
we turn on a proper \emph{$G$-symmetric} interaction on one boundary, we can
fully gap the boundary excitations in such a way that the ground state is not
degenerate.  (Such a gapping process also do not break the $G$ symmetry.) Thus
the gapping process does not leave behind any low energy degrees of freedom on
the gapped boundary.  Now we ``gauge'' the on-site symmetry $G$ in the
$(d+1)$-dimensional bulk.  The resulting system is a non-perturbative
definition of anomaly-free chiral gauge theory described by $\nu_i$
with $\sum \nu_i=0$.  Since the thickness $l$ of the $(d+1)$-dimensional bulk is
finite (although $l$ can be large so that the two boundaries are nearly
decoupled), the system actually has a $d$-dimensional space-time.  In
particular, due to the finite $l$, the gapless gauge bosons of the gauge group
$G$ only live on the $d$-dimensional boundary.

The same approach also works for fermionic systems.  We can start with a few
fermionic SPT states in $(d+1)$-dimensional space-time described by
super-cocycles $\nu_i$\cite{GW1248} that satisfy $\sum \nu_i=0$ (\ie the
combined fermion system is free of all the gauge anomalies).  If we turn on a
proper $G$-symmetric interaction on one boundary, we can fully gap the boundary
excitations in such a way that the ground state is not degenerate and does
break the symmetry $G$.  In this case, if we gauge the bulk on-site symmetry,
we will get a non-perturbative definition of anomaly-free fermionic chiral
gauge theory.

\noindent
\textbf{A non-perturbative definition of an $SO(10)$ chiral gauge theory}: To
define an $SO(10)$ chiral gauge theory\cite{FM7593} in $4$-dimensional
space-time, we start with a free fermion hopping model on a $4$-dimensional
space lattice (with a continuous time).  We design the free fermion hopping
model such that there is a fermion band gap in the bulk and there is a single
two-component gapless Weyl fermion mode on the boundary (see 
%Supplemental Material at [URL will be inserted by publisher] 
appendix A
for a particular construction).\cite{CH9497,QHZ0824}
We also assume that the $4$-dimensional space lattice form a slab of thickness
$l$.  The  massless  Weyl fermions on one boundary is described by
the following Hamiltonian
$ H=- \psi^\dag \ii \si^i \prt_i \psi$,
where $\psi$ is a two-component Weyl fermion operator, and $\si^l$, $l=1,2,3$
are the Pauli matrices.  We will call $\psi$ the right-hand Weyl fermions.
The  massless  Weyl fermions on the other boundary is described by
left-hand Weyl fermions with a Hamiltonian
$  H=- {\t\psi}^\dag \ii (\si^i)^* \prt_i {\t\psi}$ .

Next, we take 16 copies of the above theory, which will lead to 16 gapless
right-hand Weyl fermions on one boundary
\begin{align}
 H=- \psi_\al^\dag \ii \si^i \prt_i \psi_\al,\ \ \
\al =1,\cdots,16 .
\end{align}
and 16 gapless left-hand Weyl fermions on the other boundary.  Such 16 fermions
will form the 16-dimensional spinor representation of $SO(10)$.  We note that,
by construction, the free fermion hopping model on the $4$-dimensional space
lattice has the  $SO(10)$ symmetry, which is an on-site symmetry.

Then, we add an $SO(10)$ symmetric interaction between the left-hand Weyl
fermions 
%described by \eq{Hl} 
on one boundary. If the interaction can fully gap
out the left-hand Weyl fermions (\ie give all the  left-hand Weyl fermions a
finite mass) without breaking the $SO(10)$ symmetry, then, the only low energy
excitations are the massless right-hand Weyl fermions that form the spinor
representation of $SO(10)$.  Since $l$ is finite, we can view the
$4$-dimensional slab as a  3-dimensional lattice.  Thus, we obtain a lattice
model of interacting fermions in 3-dimensional space, such that the low energy
excitations of the model are the  right-hand Weyl fermions forming the spinor
representation of $SO(10)$. The lattice model also has the $SO(10)$ on-site
symmetry.  After gauging the  $SO(10)$ on-site symmetry in 4+1D lattice theory,
we obtain a non-perturbative definition of $SO(10)$ chiral gauge theory in
terms of a lattice gauge theory in  3-dimensional space.

The key step in the above construction is to add a proper interaction between
the left-hand Weyl fermions on one boundary to gap out all the left-hand Weyl
fermions without breaking the $SO(10)$ symmetry.  Is this possible? If the
$SO(10)$ chiral fermion theory (with right-hand Weyl fermion in 16-dimensional
representation of $SO(10)$) is free of all the gauge anomalies, then almost by
definition, there will exist a proper interaction between the Weyl
fermions on one boundary to gap out all the Weyl fermions without
breaking the $SO(10)$ symmetry.  We know that the $SO(10)$ chiral fermion
theory is free of all ABJ gauge anomalies and free of all gravitational
anomalies (since the chiral fermion can all be gapped if we break the $SO(10)$
symmetry).  However, we do not know if the  $SO(10)$ chiral fermion theory is
free of all potential nonABJ anomalies (such as global gauge anomalies).  In
the following, we will propose a way to design the interaction between the Weyl
fermions so that the interaction can gap out all the Weyl fermions on one
boundary without breaking the $SO(10)$ symmetry. This suggests that the
$SO(10)$ chiral fermion theory is free of all gauge anomalies.

One way to obtain such an interaction is to introduce real scaler fields
$\phi^a$, $a=1,\cdots,10$, in the 10-dimensional representation of $SO(10)$ and
construct the following interacting theory
\begin{align}
 H&=- \t\psi_\al^\dag \ii (\si^i)^* \prt_i \t\psi_\al
+H(\phi^a)
+\t\psi^T \eps C\ga_a  \phi^a \t\psi+h.c.
\end{align}
where $\eps=\ii\si^2$ acting on the Weyl spinor index.  Here $H(\phi^a)$ is the
Hamiltonian for the scaler fields $\phi^a$, and the 16-by-16 matrices $C$ and
$\ga_a$ are chosen such that $\t\psi^T \eps C  \phi^a \t\psi$ form the
10-dimensional representation of $SO(10)$ (see 
%Supplemental Material at [URL will be inserted by publisher] 
appendix B
for details).\cite{Zee03} $ C  \ga_a \phi^a$ can
be viewed as a hermitian matrix with eight eigenvalues equal to
$\sqrt{\phi^a\phi^a}$  and eight eigenvalues equal to $-\sqrt{\phi^a\phi^a}$.
Therefore, the term $\t\psi^T \eps C\ga_a  \phi^a \t\psi+h.c.$ generate a mass
$M=\sqrt{\phi^a\phi^a}$ for all the 16 Weyl fermions if the $\phi^a$ field is a
non-zero constant.  The  non-zero constant $\phi^a$ field break the $SO(10)$
symmetry. The fact that the 16 Weyl fermions can be fully gapped implies that
they are free of gravitational anomalies.

The Hamiltonian $H(\phi^a)$ for the real scaler field is chosen to make
$\phi^a\phi^a=M^2\neq 0$ without breaking the $SO(10)$ symmetry $\<\phi^a\>=0$.
So the orientation of the $\phi^a$ field can fluctuate freely within a sphere
$S_9$ in 10-dimensional space.  We also assume that the correlation length
$\xi$ of the $\phi^a$ field is much larger than the lattice constant.  In this
case, we expect the term $\t\psi^T \eps C\ga_a  \phi^a \t\psi+h.c.$ generate a
mass $M\sim \sqrt{\phi^a\phi^a}$ for all the 16 Weyl fermions even when the
$\phi^a$ field is fluctuating and  $\<\phi^a\>=0$.

However, the above argument may fail if the fluctuating  $\phi^a$ field in
4-dimensional space-time contains defects where $ \phi^a=0$.  Those defects
with $ \phi^a=0$ can give rise to massless (or gapless) fermionic excitations.
Point-defect in space-time with $ \phi^a=0$ (such as instantons) can exist if
$\pi_3(S_9)\neq 0$, line-defect in space-time with $\phi^a=0$ (such as
``hedgehog'' solitons) can exist if $\pi_2(S_9)\neq 0$, membrane-defect in
space-time with $ \phi^a=0$ (such as vortex lines) can exist if $\pi_1(S_9)\neq
0$, 3D-brane-defect in space-time with $ \phi^a=0$ (such as domain walls) can
exist if $\pi_0(S_9)\neq 0$.  However, $\pi_d(S_9)=0$ for $0\leq d<9$.  So
there are no defects with $ \phi^a=0$.  We may assume the fluctuating  $\phi^a$
field satisfying $\phi^a\neq 0$ anywhere in space-time.

The  above argument may also fail if the effective Lagrangian for the
non-vanishing fluctuating  $\phi^a$ field in 4-dimensional space-time contains
a Wess-Zumino-Witten (WZW) term (the WZW term can be well-defined for
non-vanishing  $\phi^a$ field),\cite{WZ7195,W8322} after we integrating
out the massive fermions in the 4+1D bulk.  In this case, $\phi^a$ field may
not have a gapped phase that do not break the symmetry, as discussed in
\Ref{W8322,CLW1141,CGL1172,CGL1204}.  However, since $\pi_5(S_9)=0$, the
non-vanishing $\phi^a$ field in 4-dimensional space-time cannot have any WZW
term.

The above considerations make us to believe that the term $\t\psi^T \eps C\ga_a
\phi^a \t\psi+h.c.$ does generate a mass $M\sim \sqrt{\phi^a\phi^a}$ for all
the 16 Weyl fermions even when $\<\phi^a\>=0$ and the $SO(10)$ symmetry is not
broken.  The fact that the 16 Weyl fermions can be fully gapped without
breaking the $SO(10)$ symmetry implies that they are free of \emph{all}
$SO(10)$ gauge anomalies.\cite{W1303}

The above argument can be generalized to other symmetries, which leads to the
conjecture stated at the begining of the paper.  In the above $SO(10)$ example,
the symmetry breaking fields $\phi^a$ can generate the (Higgs) mass terms in
the conjecture that give all the fermions a mass gap. The  unbroken symmetry
group $G_\text{grnd}$ in the conjecture is $SO(9)$.  The configurations of the
symmetry breaking fields generated by the $SO(10)$ rotations forms a space
$G/G_\text{grnd}=SO(10)/SO(9)=S_9$.

Next, we will try to apply our anomaly-free conditions to some other chiral
fermion theories.  If the two conditions are satisfied, then the  chiral fermion
theory is free of all anomalies.  If not, the theory may or may not have
anomalies.  For a chiral fermion theory with $U(1)$ gauge symmetry, any mass
term will break the $U(1)$ symmetry, and thus $G_\text{grnd}=1$ (\ie trivial).
We have $\pi_1(G/G_\text{grnd})=\Z$, and the condition (2) is not satisfied.
So the theory can be anomalous which is a correct result.  Next, let us
consider a chiral fermion theory with a $SU(2)$ gauge symmetry.  The theory
contains two right-hand fermions forming an $SU(2)$ doublet.  The theory also
contains two left-hand fermions which are $SU(2)$ singlet.  We can make all the
fermions massive by breaking the $SU(2)$ symmetry completely (\ie
$G_\text{grnd}=1$).  Since $\pi_3(G/G_\text{grnd})=\pi_3[SU(2)]=\Z$, the the
condition (2) is not satisfied for 2-dimensional space-time and above. So the
theory can be anomalous in 2-dimensional space-time and above, which is again
correct.  The above two examples demonstrate that our argument does not apply
for known anomalous theories.

\noindent \textbf{Summary}: In this paper, we proposed a way to construct a
lattice gauge model to non-perturbatively define a 3+1D $SO(10)$ chiral gauge
theory with two-component massless Weyl fermions in the 16-dimensional spinor
representation of $SO(10)$.  The close connection between gauge anomalies and
the SPT orders allows us to show that any chiral gauge theory can be
non-perturbatively defined by putting it on a lattice  of the same dimension,
as long as the chiral gauge theory is free of all anomalies.  Such construction
is achieved by adding a proper strong interaction among the fermions. As a key
result, we propose a general way to add/design such an interaction.

The  3+1D $SO(10)$ chiral gauge theory on lattice can be combined with Higgs
fields to break the $SO(10)$ gauge ``symmetry'' to $U(1)\times SU(2)\times
SU(3)$ gauge ``symmetry'', which leads to the modified standard model and its
non-perturbative definition on lattice.  Such a procedure was studied under the
$SO(10)$ grand unified theory.\cite{FM7593}

I would like to thank Micheal Levin, Natalia Toro, and Neil Turok for helpful
discussions.  This research is supported by NSF Grant No.  DMR-1005541, NSFC
11074140, and NSFC 11274192.  It is also supported by the BMO Financial Group
and the John Templeton Foundation.  Research at Perimeter Institute is
supported by the Government of Canada through Industry Canada and by the
Province of Ontario through the Ministry of Research.

\vfill\eject

\bibliography{../../bib/wencross,../../bib/all,../../bib/publst,./tmp}

\appendix \section{The lattice model} \label{latt}

The lattice model in 4D space, whose boundary gives rise to a single massless
Weyl fermion, has the following form
\begin{align}
 H=H_\text{hop}+H_\text{int},
\end{align}
where
\begin{align}
 H_\text{hop}=\sum_{ij} (t_{ij}^{\al\bt} c^\dag_{\al,i} c_{\bt,j} +h.c.)
\end{align}
is a lattice fermion hopping model with $16\times 4$ fermion orbitals (labled by
$\al,\bt=1,\cdots ,16\times 4$) per site.  $H_\text{int}$ describe the interaction between
the fermions.

Let us first construct
\begin{align}
 H^1_\text{hop}=\sum_{ij} (t_{ij}^{ab} c^\dag_{a,i} c_{b,j} +h.c.)
\end{align}
which has 4 fermion orbital per site ($a,b=1,\cdots ,4$).  
To construct $H^1_\text{hop}$, let us
introduce
\begin{align}
 \Ga^1&=\si^1\otimes \si^3,\ \ \
 \Ga^2=\si^2\otimes \si^3,\ \ \
\\
 \Ga^3&=\si^0\otimes \si^1,\ \ \
 \Ga^4=\si^0\otimes \si^2,\ \ \
 \Ga^5=\si^3\otimes \si^3 ,
\nonumber
\end{align}
which satisfy
\begin{align}
 \{\Ga^i,\Ga^j\}=2\del_{ij} .
\end{align}
In the $k$ space, the lattice model $H_\text{hop}^1$ is given by the following
one-body Hamiltonian
\begin{align}
\label{H4D}
&\ \ \ \
 H( k_1, k_2, k_3, k_4)
\\
&=
2[\Ga^1\sin(k_1) +\Ga^2\sin(k_2) +\Ga^3\sin(k_3) +\Ga^4\sin(k_4)]
\nonumber\\
&\ \ \ \
+2\Ga^5[\cos(k_1) +\cos(k_2) +\cos(k_3) +\cos(k_4)-3]
.
\nonumber
\end{align}
% 0000  4 + \\
% 1000  2 - \\
% 0100  2 - \\
% 1100  2 + \\
% 0010  2 - \\
% 1010  0 + \\
% 0110  0 + \\
% 1110 -2 - \\
% 0001  2 - \\
% 1001  0 + \\
% 0101  0 + \\
% 1101 -2 - \\
% 0011  0 + \\
% 1011 -2 - \\
% 0111 -2 - \\
% 1111 -4 + \\
Since the band structure of such a 4D hopping model \eq{H4D} is designed to
have a non-trivial twist, the 4D lattice model will have one two-component
massless Weyl fermion on its 3-dimensional surface, appearing at the zero
energy (single-body energy).\cite{CH9497,QHZ0824}

Let us consider a 4-dimensional lattice formed by stacking two 3-dimensional
cubic lattices. We then put the above 4-dimensional lattice fermion hopping
model on such a 4-dimensional lattice which has only two layers in the
$x^4$-direction.  The one-body Hamiltonian in the $(k_1,k_2,k_3)$-space is
given by the following 8-by-8 matrix
\begin{align}
 H(k_1,k_2,k_3)
= \begin{pmatrix}
M_1 & M_2\\
M_2^\dag & M_1
\end{pmatrix}
\end{align}
where
\begin{align}
M_1 &= 2[\Ga^1 \sin(k_1) +\Ga^2 \sin(k_2) +\Ga^3 \sin(k_3)]
\nonumber\\
& \ \ \ \ \ \
+2\Ga^5 [ \cos(k_1) +\cos(k_2) +\cos(k_3) -3],
\nonumber\\
M_2 &= -\ii \Ga^4 +  \Ga^5
.
\end{align}
We find that the above fermion hopping model give rise to one two-component
massless Weyl fermion on each of the two surfaces of the 4D lattice.  (A
surface is a 3D cubic lattice.) The  Weyl fermion on one boundary is left-hand
Weyl fermion and the  Weyl fermion on the other boundary is right-hand Weyl
fermion.

The above hopping model is defined on a 4D lattice with only two layers of 3D
cubic lattices.  We may also construct a hopping model on a 4D lattice with $l$
layers.  In this case, we still get one two-component Weyl fermion on each of
the two surfaces of the 4D lattice.  However, the  two-component Weyl fermions
on different surfaces has a mixing of order $\ee^{-l}$, which gives the fermion
a Dirac mass of order  $\ee^{-l}$.  (Our two-layer model is fine tuned to make
such a mixing vanishes.)

Then we put 16 copies of the above hopping model $H^1_\text{hop}$ together to
obtain a hopping model $H_\text{hop}$ with an $SO(10)$ symmetry (where fermions
form the 16-dimensional spinor representation of $SO(10)$).  Next we try to
include a proper $SO(10)$ symmetric interaction among fermions on only one
boundary to give those, say left-hand, fermion a mass term of order cut-off
scale without breaking the $SO(10)$ symmetry.  In the main text, we discussed
how to design such an interaction [via scalar fields $\phi^a$ in the
10-dimensional representation of $SO(10)$].  Since the target space of the
scalar fields $\phi^a$ is $S_9$ which has trivial homopoty group $\pi_d(S_9)=0$
for $d<9$, we argue that such a  scalar field can generate an interaction term
$H_\text{int}$ which gives the left-hand fermion on one boundary a mass term of
order cut-off scale without breaking the $SO(10)$ symmetry.

Since the mixing of the fermions on
the two boundaries is of order $\ee^{-l}$, the interaction on one boundary will
only induce a weak $SO(10)$ symmetric interaction of order $\ee^{-l}$ on the
other boundary.  Since all the interactions are irrelavent, any weak
interactions cannot give the right-hand fermions on the other boundary a mass
term.  The right-hand fermions on the other boundary will be massless.

Once we put the right-hand Weyl fermions on lattice with the full $SO(10)$
symmetry (realized as an on-site symmetry), then it is easy to gauge the global
(on-site) $SO(10)$ symmetry to obtain a lattice $SO(10)$ gauge model which
produces  right-hand massless Weyl fermions coupled to $SO(10)$ gauge field, at
low energies.

\section{$SO(10)$ spinor representations}
\label{higgs}

To understand the $SO(10)$ spinor representations,\cite{Zee03} let us introduce
$\ga$-matrices $\ga_a$, $a=1,\cdots,10$:
\begin{align}
 \ga_{2k-1}&=
\underbrace{\si^0\otimes \cdots \otimes \si^0}_{k-1 \ \si^0{\text{'s}}}
\otimes \si^1\otimes
\underbrace{\si^3\otimes \cdots \otimes \si^3}_{5-k \ \si^3{\text{'s}}}
\nonumber\\
 \ga_{2k}&=
\underbrace{\si^0\otimes \cdots \otimes \si^0}_{k-1 \ \si^0{\text{'s}}}
\otimes \si^2\otimes
\underbrace{\si^3\otimes \cdots \otimes \si^3}_{5-k \ \si^3{\text{'s}}}
\nonumber\\
k&=1,\cdots,5 ,
\end{align}
which satisfy
\begin{align}
 \{\ga_a,\ga_b\}=2\del_{ab},\ \ \ \ \ga_a^\dag=\ga_a.
\end{align}
Here $\si^0$ is the 2-by-2 identity matrix and $\si^l$, $l=1,2,3$ are the Pauli
matrices.  The 45 hermitian matrices
\begin{align}
 \Ga_{ab}=\frac{\ii}{2}[\ga_a,\ga_b]=
\ii \ga_a\ga_b, \ \ \ \ a<b,
\end{align}
generate a 32-dimensional representation of $SO(10)$:
$\ee^{\ii \th^{ab} \Ga_{ab}}$, $\th_{ab}=-\th_{ba}$.
The above 32-dimensional representation is reducible.
To obtain irreducible representation, we introduce
\begin{align}
\ga_\text{FIVE}&=(-)^5 \ga_1\otimes\cdots\otimes\ga_{10}
=
\underbrace{\si^3\otimes \cdots \otimes \si^3}_{5 \ \si^3{\text{'s}}},
\nonumber\\
(\ga_\text{FIVE})^2&=1,\ \ \ \
\Tr \ga_\text{FIVE}=0.
\end{align}
We see that
$\{\ga_\text{FIVE},\ga_a\}=[\ga_\text{FIVE},\Ga_{ab}]=0$.
This allows us to obtain two 16-dimensional irreducible representations
\begin{align}
&
\ee^{\ii \th^{ab} \Ga^+_{ab}}:\  \Ga^+_{ab}=
\frac{1+\ga_\text{FIVE}}{2} \Ga_{ab} \frac{1+\ga_\text{FIVE}}{2} ,
\nonumber\\
&
\ee^{\ii \th^{ab} \Ga^-_{ab}}:\  \Ga^-_{ab}=
\frac{1-\ga_\text{FIVE}}{2} \Ga_{ab} \frac{1-\ga_\text{FIVE}}{2} .
\end{align}

The two 16-dimensional irreducible representations are related.
Let us introduce
\begin{align}
 C=
\si^2\otimes\si^1\otimes
\si^2\otimes\si^1\otimes
\si^2,
\end{align}
which satisfies
\begin{align}
 C^{-1}\Ga^*_{ab}C&=-\Ga_{ab},\ \ \ \ \
 C^{-1}\ga^*_aC=-\ga_a,
\nonumber\\
 C^{-1}\ga_\text{FIVE}C&=-\ga_\text{FIVE}
.
\end{align}
If the Weyl fermion operators $\psi_+$ form the 16-dimensional irreducible
representation $\Ga^+_{ab}$, then $\psi_-=C \psi_+^*$ is the other
16-dimensional irreducible representation $\Ga^-_{ab}$.

Using the above results, we can show that $\psi_+^T \eps C\ga_a \psi_+$ form a
10-dimensional representation of $SO(10)$,
since
\begin{align}
 [\Ga_{ab},\ga_c]=-2\ii (\del_{ac}\ga_b-\del_{bc}\ga_a).
\end{align}
The above leads to
\begin{align}
&\ \ \
\psi_+^T \eps C\ga_a \psi_+=
\psi_+^T \eps \frac{1+\ga_\text{FIVE}}{2}C\ga_a\frac{1+\ga_\text{FIVE}}{2} \psi_+
\nonumber\\
& \to
\psi_+^T \eps \ee^{\ii \th^{ab} \Ga_{ab}^T}\frac{1+\ga_\text{FIVE}}{2} C\ga_a\frac{1+\ga_\text{FIVE}}{2} \ee^{\ii \th^{ab} \Ga_{ab}} \psi_+
\nonumber\\
& =
\psi_+^T\eps \frac{1+\ga_\text{FIVE}}{2} C C^{-1}\ee^{\ii \th^{ab} \Ga_{ab}^*} C\ga_a \ee^{\ii \th^{ab} \Ga_{ab}} \frac{1+\ga_\text{FIVE}}{2}\psi_+
\nonumber\\
& =
\psi_+^T\eps \frac{1+\ga_\text{FIVE}}{2} C \ee^{-\ii \th^{ab} \Ga_{ab}} \ga_a \ee^{\ii \th^{ab} \Ga_{ab}} \frac{1+\ga_\text{FIVE}}{2}\psi_+
\nonumber\\
&= G_a^b (\th_{ab})
\psi_+^T\eps \frac{1+\ga_\text{FIVE}}{2} C  \ga_b \frac{1+\ga_\text{FIVE}}{2}\psi_+,
\nonumber\\
&= G_a^b (\th_{ab})
\psi_+^T \eps C  \ga_b \psi_+,
\end{align}
where the 10-by-10 matrix $G(\th_{ab}) \in SO(10)$.  Here, we may view $C
\ga_b$ and $\Ga_{ab}$ as 16-by-16 matrices acting within the 16-dimensional
space with $ \frac{1+\ga_\text{FIVE}}{2}=1$.  Note that $C  \ga_b$ and
$\Ga_{ab}$ commute with  $ \frac{1+\ga_\text{FIVE}}{2}$.  When viewed as such a
16-by-16 matrix, $C  \ga_{10}$ is a real symmetric matrix with eight
eigenvalues equal to 1 and eight eigenvalues equal to $-1$.

\section{A more detailed discussion on mirror fermion approach}
\label{EP}

The approach propose in this paper is similar to the mirror fermion proposed in
\Ref{EP8679,M9259,BMP0628,GP0776}.  Both approaches try to solve the chiral
fermion problems via the usual lattice gauge theory by simply adding direct
fermion-fermion interactions to gap out the unwanted mirror  sector.

Certainly, not every chiral fermion theory can be defined on a lattice using
such approaches.  The main difference between the two approaches is in the proposed
conditions for the mirror sector to be fully gappable without breaking
the required gauge symmetry.  In this paper, we propose a rather conservative
sufficient condition (slightly generalized):\\
\frm{\textbf{Statement A:} \emph{A chiral fermion theory in $d$-dimensional
space-time with a gauge group $G_g$ can be defined on a lattice if
(0) the free  chiral fermion theory without mass term
has a symmetry $G$ which may be equal to or bigger than $G_g$;
(1) there exist (possibly symmetry breaking) mass terms that make
all the fermions massive; and (2) $\pi_n(G/G_\text{grnd})=0$ for $n\leq d+1$,
where $G_\text{grnd}$ is the unbroken symmetry group.}}
In \Ref{EP8679} it was
stated that ``\emph{Elementary fermions transforming as a complex
representation of the gauge group are able to acquire explicit masses
consistent with the gauge symmetry by pairing up with composite fermion states
transforming as the conjugate representation of the gauge group.  The composite
fermion states are bound, not by the gauge interaction, but by an auxiliary
interaction which has been introduced for this explicit purpose.}'' To compare with
our result, we interpret the above as:
\frm{\textbf{Statement B:} \emph{A chiral fermion theory with a
gauge group $G$ can be defined on a lattice if there exist composite
fermion operators formed by mirror fermions, such that there are gauge
invariant mass terms between  composite fermion operators and the
mirror fermion operator to fully gap out the composite fermions and the
mirror fermions.}}

Let us apply the Statement B to the 3-4-5-0 model in 1+1D, with two
right-moving mirror fermions $\psi_3$ and $\psi_4$ of $U(1)$ charge 3 and 4 and
two left-moving mirror fermions $\bar\psi_5$ and $\bar\psi_0$ of $U(1)$ charge
$-5$ and 0.  The composite fermions are
\begin{align}
\bar\chi_3 &= \bar\psi_0 (\psi_3 \bar\psi_5) (\psi_4 \bar\psi_5),
	&	
\bar\chi_4 &= \bar\psi_0 (\psi_3 \bar\psi_5)^2 ,
	\nonumber\\
\chi_5 &= \psi_4 (\psi_4^C \bar\psi_5^C),
	&
\chi_0 &= \psi_3 (\psi_3 \bar\psi_5)(\psi_4 \bar\psi_5).
\end{align}
where $C$ is the charge conjugation which maps right-movers (left-movers) to  right-movers (left-movers).
The composite fermions and the mirror fermions can be fully gapped by the mass
term $ \bar\chi_3\psi_3 + \bar\chi_4\psi_4 + \bar\psi_5\chi_5 +
\bar\psi_0\chi_0$, and the statement B implies that the  3-4-5-0 model in 1+1D
can be defined on lattice. This result agrees with \Ref{WW1380} but is not
supported by \Ref{CGP1247}.

We can also apply the Statement B to the 3-4-5-2 model in 1+1D, with two
right-moving mirror fermions $\psi_3$ and $\psi_4$ of $U(1)$ charge 3 and 4 and
two left-moving mirror fermions $\bar\psi_5$ and $\bar\psi_2$ of $U(1)$ charge
$-5$ and $-2$.  The composite fermions are
\begin{align}
\bar\chi_3 &= \bar\psi_2 (\psi_4 \bar\psi_5),
	&
\bar\chi_4 &= \bar\psi_2 (\psi_3 \bar\psi_5) ,
	\nonumber\\
\chi_5 &= \psi_4 (\psi_4^C \bar\psi_5^C),
	&
\chi_2 &= \psi_3 (\psi_4^C \bar\psi_5^C).
\end{align}
The composite fermions and the mirror fermions can be fully gapped by the mass
term $ \bar\chi_3\psi_3 + \bar\chi_4\psi_4 + \bar\psi_5\chi_5 +
\bar\psi_2\chi_2$, and the statement B implies that the  3-4-5-2 model in 1+1D
can be defined on lattice. This result is incorrect since the  3-4-5-2 model in
1+1D has an $U(1)$ gauge anomaly.

So our interpretation of the result in \Ref{EP8679}, the Statement B, is
incorrect.  But it is not clear what is the general result from
\Ref{EP8679,M9259,BMP0628,GP0776} to compare with our general result Statement
A.

\end{document}